\documentclass[a4paper,8pt]{article}

\begin{document}

\newcommand{\munu}{^\mu_{\phantom{\mu}\nu}}
\makeatletter
\@addtoreset{equation}{section}
\def\theequation{\thesection.\arabic{equation}}
\makeatother

\title{Additional information decreases the estimated entanglement using the Jaynes principle}
\author{Koji Nagata (nagata@kaist.ac.kr)\\
{\it Department of Physics,}\\
{\it Korea Advanced Institute of Science and 
Technology,}\\
{\it Daejeon 305-701, Korea}}

\date{}
\maketitle

\begin{abstract}
We study a particular example considered in
{[Phys. Rev. A {\bf 59,} 1799 (1999)]}, concerning the statistical inference of quantum 
entanglement using the Jaynes principle. 
Assume a Clauser-Horne-Simony-Holt (CHSH) Bell operator, 
a sum of two operators $\sqrt{2}(X+Z)$.
Given only an average of the Bell-CHSH operator, 
we may overestimate entanglement.
However, 
the estimated entanglement is decreased (never increases) when we use the 
expectation value of the operator $X$ as additional information. 
A minimum entanglement state is obtained by
minimizing the variance of the observable $X$.
\end{abstract}

{\it Keywords:} 
Entanglement, the Jaynes principle.


\newpage

\section{Introduction}

The importance of understanding quantum entanglement in the field of quantum information theory \cite{NC,Galindo} has been recognized.
Especially, concerning what kind of entangled states are possible
under constraints of incomplete experimental data, 
there is the problem of determining how much entanglement is likely
under the constraints. 
Such problems of estimation for bipartite entanglement have been discussed 
in the context of the maximum entropy principle 
 (the Jaynes principle) \cite{bib:Jaynes}.
In 1999, Horodecki {\it et al.} invoked \cite{bib:Horodecki} the Jaynes principle for statistical inference of incomplete data (a single observable)
in the quest for quantifying entanglement.

The Jaynes principle implies that if some number of expectation values of an incomplete but linearly independent set of observables has been measured, the state of the system is determined by maximizing the von Neumann entropy subject to the given constraints, along with the normalization of the density matrix.

Application of this principle sometimes leads to an estimated state 
that possesses stronger entanglement 
than the minimum entanglement that is compatible
with the measured data.
Horodecki {\it et al.} suggested an inference scheme which should not give us an inseparable estimated state if theoretically there is a separable state compatible with the measured data.
They claimed further that in all the quantum information processes where entanglement is needed, the proper inference scheme should involve minimization of entanglement. 
Therefore, they introduced a new constraint, i.e., minimization of
entanglement in applying the maximum entropy principle.
They thus obtained an estimated state that has the minimum entanglement.
Rajagopal derived the same state with a different assumption together with
the maximum entropy principle, i.e., minimizing the variance of a Bell
operator \cite{bib:Rajagopal}.
Since then,
considerable attention has been paid \cite{1,2,3,4,5,6,7,8} to this problem.

Slater in his 2006 paper \cite{8} makes this most clear of all the papers
extant on the issue: Jaynes' principle simply gives us a family of states
for any set of functions of the density operator (not including the
entropy, obviously). See also the point \cite{3} expressed by Brun {\it et al}.
They expressed that, if there is no specific 
additional information, the maximum entropy state is preferable to the minimum entanglement
state. They invoked a projection measurement in the Bell basis 
(cf. Eq.~(\ref{Bell basis})).
They further considered how to turn a single Jaynes state into 
a maximally entangled state.
Finally, they concluded that the prediction of maximum entropy state 
causes no practical difficulty. It merely represents the fact that,
in some sense, most states that are consistent with the constraint are also 
entangled.

Here, we investigate a particular example considered in \cite{bib:Horodecki}
from the point of view of the statistical inference of quantum 
entanglement using the Jaynes principle. 
Assume a Clauser-Horne-Simony-Holt (CHSH) \cite{CHSH} Bell operator, 
a sum of two operators $\sqrt{2}(X+Z)$.
Given an average of only the Bell-CHSH operator $\sqrt{2}\langle X+Z \rangle$,
we may overestimate entanglement.
However, 
the estimated entanglement is 
decreased when we have the expectation value of one of the operators, 
$\langle X\rangle$,
constituting the average value of
the Bell-CHSH operator as further
knowledge.
We indicate that a
minimum entanglement state is also derived within the philosophy espoused by
Rajagopal, that is, 
minimizing 
the variance of the observable $X$.
We hope the discussion presented here contribute to the concept of entanglement and entropy, in regard to the amount of information.

\section{Decrease of overestimated entanglement}
In what follows, we assume that the average value of 
the following Bell-CHSH operator is given as 
in Ref.~\cite{bib:Horodecki}
\begin{eqnarray}
2\sqrt{2}B=\sqrt{2}(\sigma_x^1\sigma_x^2+\sigma_z^1\sigma_z^2)=\sqrt{2}(X+Z)
\end{eqnarray}
where 
\begin{eqnarray}
B=\frac{\sigma_x^1\sigma_x^2+\sigma_z^1\sigma_z^2}{2},~
X=\sigma_x^1\sigma_x^2,~
Z=\sigma_z^1\sigma_z^2.
\end{eqnarray}
This is equivalent to the expectation value $\langle B\rangle$
being given.
We suppose that $\langle B\rangle\geq 1/2$. 
The estimated state is an entangled state if 
$\langle B\rangle> 1/2$ \cite{5}.
We use the Bell basis as in Ref.~\cite{bib:Horodecki}
\begin{eqnarray}
|\Phi^{\mp}\rangle=\frac{1}{\sqrt{2}}
(|\uparrow\uparrow\rangle\mp|\downarrow\downarrow\rangle),~
|\Psi^{\pm}\rangle=\frac{1}{\sqrt{2}}
(|\uparrow\downarrow\rangle\pm|\downarrow\uparrow\rangle).\label{Bell basis}
\end{eqnarray}
It is obvious that $\langle B\rangle> 1/2$ implies that the fidelity 
distance to the Bell state $|\Phi^+\rangle$ is more than one-half, that is, the unknown state should be an
entangled state. 
On the other hand, a separable state is compatible with experimental data
$\langle B\rangle=1/2$ (cf. Eq.~(\ref{sep})).

Using the Jaynes principle with $\langle B\rangle$, we get the following estimated state:
\begin{eqnarray}
&&\rho_{J1}=
\left(\frac{1+\langle B\rangle}{2}\right)^2
|\Phi^+\rangle\langle\Phi^+|+
\left(\frac{1-\langle B\rangle}{2}\right)^2
|\Psi^-\rangle\langle\Psi^-|\nonumber\\
&&\quad+\left(\frac{1-\langle B\rangle^2}{4}\right)
(|\Psi^+\rangle\langle\Psi^+|+|\Phi^-\rangle\langle\Phi^-|).
\end{eqnarray}
Thus $\rho_{J1}$ is inseparable 
even though some separable state satisfies
the experimental data with 
$\langle B\rangle=1/2$---it might be that entanglement
was overestimated.

Assume that we obtain an expectation value $\langle X \rangle$
as additional information. 
The original information $\langle B \rangle$ is fixed.
Thus, one has the following maximum entropy state after some algebra:
\begin{eqnarray}
&&\rho_{J2}=
\left(\frac{1-\langle X\rangle^2+
2\langle B\rangle+2\langle X\rangle \langle B\rangle}{4}\right)
|\Phi^+\rangle\langle\Phi^+|\nonumber\\
&&+
\left(\frac{1-\langle X\rangle^2-2\langle B\rangle+2\langle X\rangle\langle B\rangle}{4}\right)
|\Psi^-\rangle\langle\Psi^-|\nonumber\\
&&+
\left(\frac{1+2\langle X\rangle+\langle X\rangle^2-2\langle B\rangle-2\langle X\rangle \langle B\rangle}{4}\right)
|\Psi^+\rangle\langle\Psi^+|\nonumber\\
&&+\left(\frac{1-2\langle X\rangle+\langle X\rangle^2+2\langle B\rangle-
2\langle X\rangle\langle B\rangle}{4}\right)
|\Phi^-\rangle\langle\Phi^-|.\nonumber\\
\end{eqnarray}

Now, we shall 
show that the amount of entanglement is decreased (never increases).
As discussed in Ref.~\cite{bib:Horodecki}, we shall use two measures: 
One is the entanglement of formation $E_1$ \cite{Bennett}.
Another is the relative entropy of entanglement $E_2$ \cite{Vedral}.
For Bell diagonal states, both measures 
depend only on the largest eigenvalue $f$ of the given state and increasing functions of $f$
\begin{eqnarray}
&&E_1=h\left(1/2+\sqrt{f(1-f)}\right),\nonumber\\
&&E_2=\ln 2-h(f),\nonumber\\
&&h(x)=-x\ln x-(1-x)\ln (1-x).
\end{eqnarray}
Then, we shall compare 
the largest eigenvalue of $\rho_{J1}$ with that of $\rho_{J2}$.
They are depicted by $f_{J1}$ and $f_{J2}$, respectively.
One has
\begin{eqnarray}
&&f_{J1}=\left(\frac{1+\langle B\rangle}{2}\right)^2,\nonumber\\
&&f_{J2}=\left(\frac{1+\langle B\rangle}{2}\right)^2-
\left(\frac{\langle X\rangle-\langle B\rangle}{2}\right)^2,
\end{eqnarray}
where the largest eigenvalue is $\langle \Phi^+|\rho|\Phi^+\rangle$
since $\langle B\rangle\geq 1/2$.
Thus, one has $f_{J1}\geq f_{J2}$.
This implies that
\begin{eqnarray}
E_1(\rho_{J1})\geq E_1(\rho_{J2}),~
E_2(\rho_{J1})\geq E_2(\rho_{J2}).
\end{eqnarray}
Hence, we have shown 
that the amount of entanglement is decreased if one has additional 
information such that
$\langle X\rangle\neq \langle B\rangle$.

The most interesting, and the new aspect of the present example is that
indeed the entanglement for Jaynes' state with $\langle B\rangle$ fixed, 
$\rho_{J1}$, is
always greater than or equal to the entanglement of the Jaynes state with 
$\langle B\rangle$
and $\langle X\rangle$ fixed, $\rho_{J2}$. 
This is implied (by the convexity of $E_1$ and of $E_2$), and in fact
entirely rests on the fact that $\rho_{J1}$ cannot be written as a average
of states of the form, $\rho_{J2}$, with any distribution over 
$\langle X\rangle$. In
other words, despite the fact that the Jaynes principle predicts a state
$\rho_{J1}$ with prescribed $\langle B\rangle$, it does not 
predict a consistent distribution
of $\langle X\rangle$! 
This observation has probably been made before in the context
of Jaynes-type inference.

\section{Minimum variance principle}
We shall show that the minimum variance principle of Rajagopal indicates that
$\rho_{J2}$ is a minimally entangled state.
Here, we consider the variance of the observable $X$:
\begin{eqnarray}
\sigma^2=\langle (X-\langle X\rangle)^2\rangle=1-\langle X\rangle^2.
\end{eqnarray}
(It is worth noting that the expectation value $\langle X\rangle$ is 
given as additional information.)
We can therefore see that one has the minimum variance if $\langle X\rangle=1$.
In this case,
\begin{eqnarray}
\rho_{J2}\rightarrow \langle B\rangle
|\Phi^+\rangle\langle\Phi^+|+(1-\langle B\rangle)
|\Psi^+\rangle\langle\Psi^+|.\label{sep}
\end{eqnarray}
One sees that the largest eigenvalue (i.e., $\langle B \rangle$) is equal to the one presented in Ref.~\cite{bib:Horodecki} in the minimally entangled state.
Therefore, we have shown that the Rajagopal minimum variance principle 
gives the minimum entanglement also in the case described here.

\section{Summary}

In summary, we have investigated a particular example 
considered in Ref.~\cite{bib:Horodecki}, concerning the statistical inference of quantum 
entanglement using the Jaynes principle. 
Given only an average value of the Bell-CHSH operator $\sqrt{2}(X+Z)$, 
we may overestimate entanglement.
However, 
the overestimated entanglement is decreased when we have 
an expectation value of the operator $X$ as additional information. 
A minimum entanglement state is obtained by 
minimizing the variance of the observable $X$.
This is further supporting evidence for the scheme of Rajagopal, the minimum variance principle.

\section*{Acknowledgments}

This work has been
supported by Frontier Basic Research Programs at KAIST and K.N. is
supported by the BK21 research professorship.

\end{document}